\documentclass[letterpaper, 10 pt, conference]{ieeeconf}

\IEEEoverridecommandlockouts  
\usepackage{amsmath}
\usepackage{graphicx}
\usepackage{cite}
\DeclareGraphicsExtensions{.pdf,.png,.jpg}
\usepackage{mathtools,amssymb}
\usepackage{mathrsfs}
\usepackage{caption}
\usepackage{subcaption}
\usepackage{xcolor}

\newtheorem{remark}{Remark}
\usepackage[utf8]{inputenc}
\usepackage[english]{babel}

\usepackage{xfrac}
\title{\LARGE \bf
	Generalizing Hybrid Integrator-Gain Systems Using Fractional Calculus
}

\author{ S. Ali Hosseini$^{1}$ and  Mohammad Saleh Tavazoei$^{1}$ and Luke F. van Eijk$^{2}$ and S. Hassan HosseinNia$^{3}$ % <-this % stops a space
	\thanks{$^{1}$Mohammad Saleh Tavazoei and S. Ali Hosseini are with the Electrical Engineering Department, Sharif University of Technology, Tehran, Iran
		{\tt\small Tavazoei@ee.sharif.ir, ali\_hoseini@ee.sharif.edu  }}%
	\thanks{$^{2}$Luke F. van Eijk is with ASM Pacific Technology, Beuningen, The Netherlands
		{\tt\small luke.van.eijk@asmpt.com}}%
	\thanks{$^{3}$S. Hassan HosseinNia is with the Department of Precision and Microsystems Engineering, Delft University of
		Technology, Delft, The Netherlands
		{\tt\small s.h.hosseinniakani@tudelft.nl}}%
}

\begin{document}
	\maketitle
	\thispagestyle{empty}
	\pagestyle{empty}
	
	\begin{abstract}
		
		The Hybrid Integrator-Gain System (HIGS) has recently gained a lot of attention in control of precision motion systems. HIGS is a nonlinear low pass filter/integrator with a 52$^\circ$ phase advantage over its linear counterpart. This property allows us to avoid the limitations typically associated with linear controllers, like the waterbed effect and Bode’s gain-phase relation. In this paper, we generalize HIGS by replacing the involved integer-order integrator by a fractional-order one to adapt the phase lead from 0$^\circ$ (linear low pass filter) to 52$^\circ$ (HIGS). To analyze this filter in the frequency domain, the describing function of the proposed filter, i.e., the fractional-order HIGS, is obtained using the Fourier expansion of the output signal. In addition, this generalized HIGS is implemented in a PID structure controlling a double integrator system to validate the performance of the proposed filter in the time domain, in which by changing the fractional variable from zero to one, the output varies from the response of a linear control system to a nonlinear one.
		
	\end{abstract}

	%%%%%%%%%%%%%%%%%%%%%%%%%%%%%%%%%%%%%%%%%%%%%%%%%%%%%%%%%%%%%%%%%%%%%%%%%%%%%%%%
	
	\section{Introduction}
	PID is one of the most popular controllers in industry \cite{AR4} because it is easily implemented and tuned, and has a simple structure. Despite its popularity, it is a linear controller, which is limited due to the waterbed effect \cite{Waterbed} and Bode's gain-phase relation \cite{AR88}. The solution to overcome the aforementioned limitations lies in the nonlinear control theory\cite{HIGSOVERCOME}. To this end, there have been several attempts, such as variable gain control \cite{AR5}, split-path nonlinear filters \cite{AR6}, and reset controllers \cite{AR9,Clegg,FORE1}.\\
	Among those, the reset controller has attracted a lot of attention since it can be easily implemented in a PID framework. J.C. Clegg was first to introduce the reset integrator in the late 1950s \cite{Clegg}, which others named after him: the Clegg integrator. This integrator can reset its state and thus has a nonlinear behavior. The Clegg integrator has only $-38.15^\circ$ phase lag but similar gain behavior compared to a linear integrator. This property is interesting because it breaks Bode's gain-phase relation. The phase lead reduces overshoot and improves the settling time of the system. In addition to the Clegg integrator, there have been several other reset controllers such as the first-order reset element (FORE) \cite{FORE1}, generalized FORE \cite{FORE2},\cite{GFrORE}, and Constant in Gain Lead in Phase (CgLp) element \cite{CgLp}.
	Reset systems, despite their advantages, have a severe drawback, which is the harsh jump on its output. This discontinuity is undesired as it may cause practical issues like actuator saturation in the motion control systems \cite{ExtendedHIGS}.\\
	The Hybrid Integrator-Gain System (HIGS) has recently been proposed to avoid the unwanted jumps in reset control systems. HIGS is presented as a piecewise linear system with no jump on its output \cite{enh}. In \cite{HIGSTr}, a linear bandpass filter has been compared to a hybrid integrator-gain-based bandpass filter to control an active vibration isolation system. In \cite{Remedy}, it has been shown that the overshoot inherent when using any stabilizing linear time-invariant feedback controller can be eliminated with a HIGS-based control strategy. In addition, the stability of HIGS-based control systems has been extensively discussed in \cite{ProjStab}, \cite{LOOPHIGS}. \\
	A frequency-domain analysis is preferred in the design of linear motion controllers since it allows intuitively ascertaining closed-loop performance measures. HIGS can also be analyzed in the frequency domain \cite{datadriven}. To this end, the describing function of HIGS is obtained from the Fourier expansion of its output signal. According to the describing function, HIGS acts as a linear low pass filter in gain but leads 52$^\circ$ in phase. Unlike generalized reset control systems \cite{Freset}, this phase lead is not controllable in HIGS. By changing the after reset value in reset control systems, the phase lead can be tuned, but doing the same for HIGS will cause discontinuity. Therefore, in this paper, we fill this gap in the state of art by generalizing HIGS using fractional calculus. Fractional calculus has been used for control designs like fractional-order PID \cite{podlubny1999fractional,xue2007fractional,tavazoei2012traditional} and CRONE control \cite{oustaloup1993great}. Recently, fractional-order elements have also been used within reset elements \cite{sebastian2021augmented}, \cite{karbasizadeh2021fractional}, in order to regulate the higher-order harmonics in the reset control systems.\\ The principal contribution of this paper is to design a new HIGS where the integer-order integrator is replaced by a fractional-order one. By varying the order of the integrator from one to zero, the proposed fractional-order HIGS varies between nonlinear and linear behavior. The design of this filter enables the extension of the reset-based CgLp element to the HIGS-based CgLp element. In addition, the describing function is obtained as another contribution to analyze the controller in the frequency domain.\\
	The remainder of this paper is organized as follows.
	In section II,  the background information of HIGS will be given, the describing function of HIGS will be discussed, and finally, some fundamental definitions of the fractional-order derivative will be brought. In section III, the fractional-order HIGS will be introduced and formulated in state-space representation. Then, the describing function corresponding to fractional-order HIGS will be derived. Finally, an approach to compensate the slope loss in the fractional integrator will be introduced and is named generalized HIGS. In section IV, an illustrative example including the controller design, implementation method, and results will be given.  Section V summarizes the main conclusions of the article.\vspace{-4mm}\\
	\section{Background}
	\subsection{Hybrid integrator-gain system}
\begin{figure}
		\centering
		\includegraphics[width=90mm,scale=0.42,trim=4 4 4 8,clip]{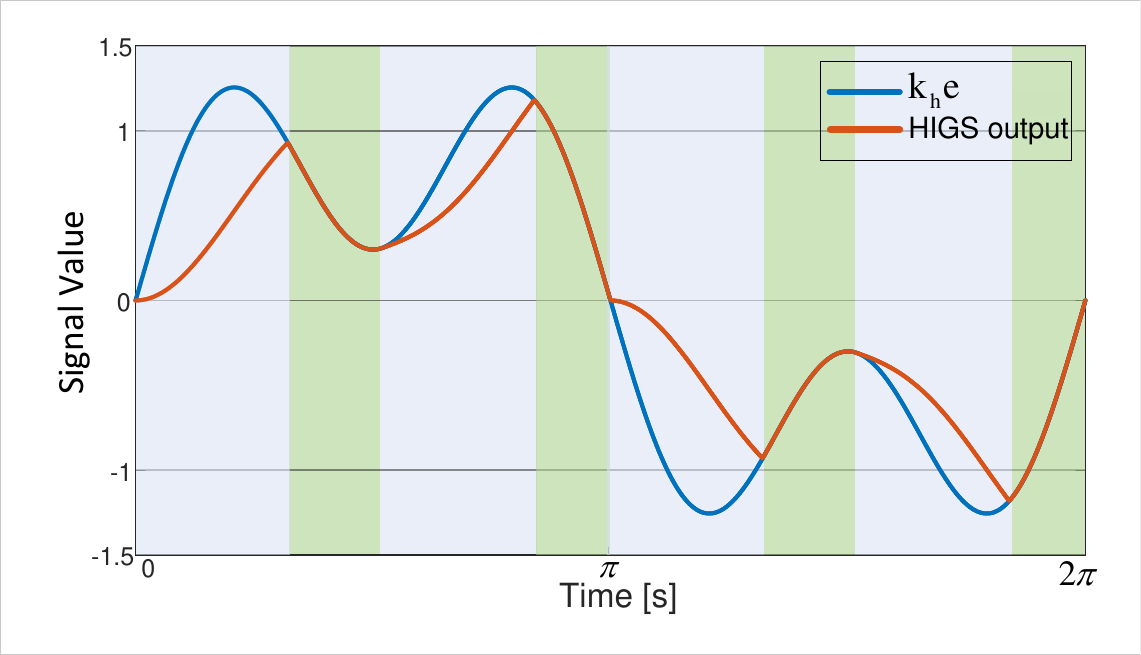}
		\setlength{\abovecaptionskip}{-5pt}
		\caption{\centering Time domain response of HIGS for multi-sine input $e(t)=\sin(t)+0.7\sin(3t)$. The blue-colored area represents integrator-mode ($F_1$), while the green-colored area represents gain-mode ($F_2$).}
		\label{fig.HIGS}
	\end{figure}

	The hybrid integrator-gain system (HIGS) is defined in \cite{enh} and its state-space representation is given by:
	\begin{equation} 
		H : \begin{cases}
			\label{eq.HIGSSS}
			\dot{x}_h=\omega_h e,       & \quad \text{if } (e,\dot{e},u) \in F_1,\\
			x_h=k_h e,       & \quad \text{if } (e,\dot{e},u) \in F_2,\\
			u=x_h,
		\end{cases}
	\end{equation}
	where $x_h\in\mathbb{R} $ is the state variable, $u\in\mathbb{R}$ is the control output, $e\in\mathbb{R} $ is the input, $ k_h \in \mathbb{R}_{>0}$ is the gain value, and $ \omega_h \in \mathbb{R}_{>0}$ is the integrator frequency. Furthermore, $F_1$ and $F_2$ denote the regions where the integrator- or gain mode is active, as defined by
% 	\begin{subequations}
% 		\label{HIGS domains}
% 		\begin{align}
% 			F_1 &:= \left\{
% 			(e,\dot{e},u) \in R^3\mid eu\geq\frac{1}{k_h}u^2 \wedge  (e,\dot{e},u) \notin F_2  \label{eq:DomF1} \right\},\\
% 			F_2 &:= \bigg\{
% 			(e,\dot{e},u) \in R^3\mid u=k_h e \wedge  \omega_h e^2 > k_h e\dot{e}  \label{eq:DomF2} \bigg\},
% 		\end{align}
% 	\end{subequations}
    \begin{subequations}
		\label{HIGS domains}
		\begin{align}
			F_1 &:= \mathcal{F} \hspace{2pt} \backslash \hspace{2pt} F_2,\\
			F_2 &:= \bigg\{
			(e,\dot{e},u) \in \mathcal{F}\mid u=k_h e \wedge  \omega_h e^2 > k_h e\dot{e}  \label{eq:DomF2} \bigg\},
		\end{align}
	\end{subequations}
	where
	\begin{equation}
        \mathcal{F} := \left\{ (e,\dot{e},u) \in \mathbb{R}^3 \hspace{2pt} | \hspace{2pt} eu \geq \frac{1}{k_h} u^2 \right\}.
    \end{equation}
	Regions $F_1$ and $F_2$ are set for three important reasons. First of all, the output stays continuous for all time. Second, the output is always bounded between $k_h e$ and zero. Finally, the control output of HIGS will always be in the same direction as the input signal, as shown in Fig. \ref{fig.HIGS}. \vspace{-4mm}\\
	\subsection{Describing function of HIGS}
\begin{figure}
		\centering
		\includegraphics[scale=0.52,trim=4 4 4 10,clip]{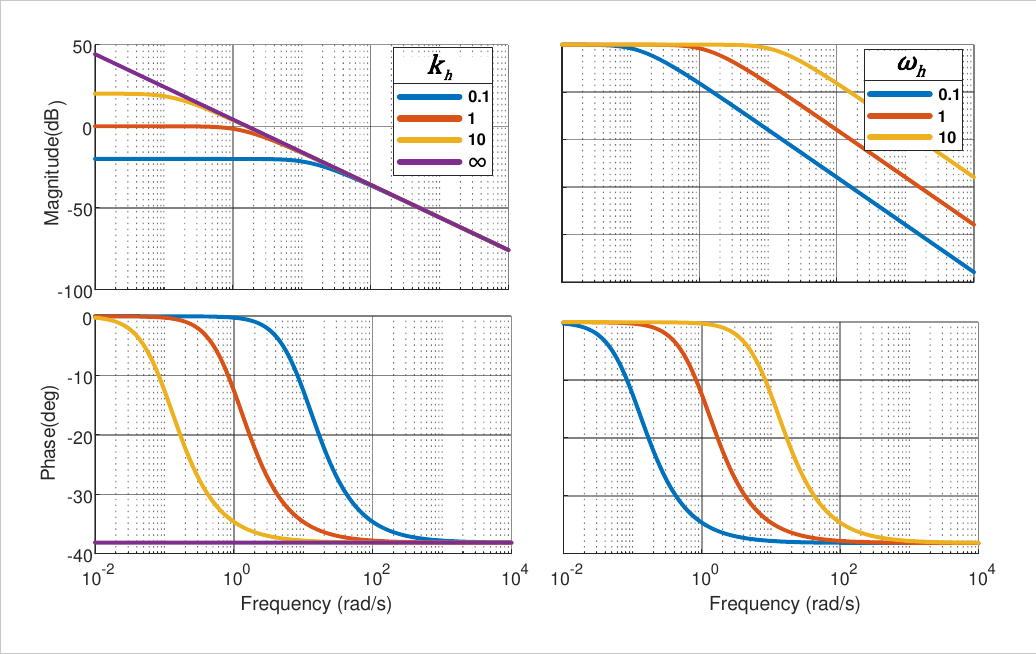}
		\setlength{\belowcaptionskip}{-15pt}
		\caption{\centering Bode plot for the describing function of HIGS for varying values of $\omega_h$ ($k_h=1$) and $k_h$ ($\omega_h=1$).}
		\label{fig.HIGS freq}
	\end{figure}

	To analyze the system in the frequency domain, a describing function analysis can be done. In \cite{datadriven} the describing function of HIGS for a sinusoidal input has been derived, as given by
	\begin{align}
		D(j\omega)=\frac{\omega_h}{j\omega}(\frac{\gamma}{\pi}+j\frac{e^{-2j\gamma}-1}{2\pi}-4j\frac{e^{-j\gamma}-1}{2\pi}) \nonumber \\ 
		+k_h(\frac{\pi-\gamma}{\pi}+\frac{e^{-2j\gamma}-1}{2\pi}) \label{HDF},
	\end{align}
	where $j := \sqrt{-1}$ is the imaginary unit, and $\gamma=2 \arctan(\frac{k_h\omega}{\omega_h})$. In Fig. \ref{fig.HIGS freq}, the Bode plot of HIGS's describing function for different values of $\omega_h$ and $k_h$ is illustrated.
	\subsection{Fractional-order derivative}
	In this section, we define the Liouville-Caputo fractional-order derivative as an approach that we will use for fractional-order HIGS calculations \cite{Frac}. Its definition is given by:
	\begin{equation}
		{}^{LC}_{}D^{\alpha}_x f(x):=\frac{1}{\Gamma(1-\alpha)}\int_ {-\infty}^x dt(x-t)^{-\alpha} \frac{df(t)}{dt},
		\label{eq:LC}
	\end{equation}
	where $0\leq \alpha \leq1$ is the derivative order, $x \in \mathbb{R}$ is the upper-bound of the integral, and $\Gamma(.)$ is the Euler Gamma function \cite{Frac}. For a sinusoidal function $f(t)$ we have \cite{exact}:
	\begin{equation} 
		\label{Frsin}
		{}^{LC}_{}D^{\alpha}_t[\sin(\omega t)]=\omega^\alpha \sin(\omega t + \frac{\pi \alpha}{2}).
	\end{equation}
	For convenience in the following, writing $LC$ in ${}^{LC}_{}D^{\alpha}_x$ will be refrained.
	\section{Generalized HIGS}
	\subsection{Fractional-order HIGS}
	HIGS is designed to dominantly operate as an integer-order integrator, except for when it tends to leave sector $\mathcal{F}$. Furthermore, HIGS has a continuous output signal. In this work, the integer-order integrator is extended to a fractional-order one. To guarantee sector-boundedness and continuity of the output signal, the gain-mode has to be revised. We define the fractional-order HIGS in state-space by
	
	\begin{equation}
        \mathcal{H}_f : \begin{cases}
        D_t^\alpha x_h = \omega_h e, & \text{if} \hspace{5pt} (e, \dot{e}, u) \in \mathcal{F}_1, \\
        x_h = k_h e, & \text{if} \hspace{5pt} (e, \dot{e}, u) \in \mathcal{F}_2, \\
        x_h = 0, & \text{if} \hspace{5pt} (e, \dot{e}, u) \in \mathcal{F}_3, \\
        u = x_h,
        \end{cases}
        \label{eq.FrHIGSSS}
    \end{equation}
    with integrator-region

    \begin{equation}
        \mathcal{F}_1 = \mathcal{F} \hspace{2pt} \backslash \hspace{2pt} (\mathcal{F}_2 \cup \mathcal{F}_3),
    \end{equation}
    upper gain-region and lower gain-region
    	\begin{subequations}
    	\begin{align}
    &\mathcal{F}_2 := \Big\{ (e,\dot{e},u) \in \mathcal{F} \hspace{2pt} | \hspace{2pt} u = k_h e \hspace{2pt} \land \hspace{2pt} \big( \omega_h D_t^{1-\alpha}(e)e > k_h \dot{e} e ... \nonumber \\
        &\lor \hspace{2pt} D_t^{1-\alpha}(e)e<0 \hspace{2pt} \lor \hspace{2pt} (e = 0 \hspace{2pt} \land \hspace{2pt} \omega_h D_t^{1-\alpha}(e)\dot{e} > k_h \dot{e}^2) \big) \Big\},
     \label{eq:zeroset2}
\end{align}
%\end{subequations}
%    	\begin{subequations}
    	\begin{align}
 &\mathcal{F}_3 = \Big\{ (e,\dot{e},u) \in \mathcal{F} \hspace{2pt} | \hspace{2pt} u = 0 \hspace{2pt} \land \hspace{2pt} \Big[ D_t^{1-\alpha}(e)e < 0 \hspace{2pt} \lor ... \nonumber \\
 &\Big(e = 0 \hspace{2pt} \land \hspace{2pt} \big( D_t^{1-\alpha}(e) \dot{e} < 0 \hspace{2pt} \lor \hspace{2pt} (\dot{e} = 0 \hspace{2pt} \land \hspace{2pt} D_t^{1-\alpha}(e) \neq 0) \big) \Big) \Big] \Big\},
   \end{align}
\end{subequations} respectively.
    The upper gain-region contains an extended version of the condition in $F_2$, which prevents violation of the sector-boundary $u = k_h e$ at $e \neq 0$ for the fractional-order integrator. The upper-gain region also contains an additional condition to prevent violation of the sector-boundary at $u = k_h e$ for $e = 0$. Namely, this can happen at that location if $\dot{e} > 0$ and $\dot{u} > k_h \dot{e}$, or if $\dot{e} < 0$ and $\dot{u} < k_h \dot{e}$. The lower gain-region contains several conditions which prevent the fractional-order HIGS from violating the sector-boundary at $u = 0$. This can happen in many different situations. For $e>0$ it happens when $\dot{u} < 0$, and for $e < 0$ when $\dot{u} > 0$. If $e = 0$, it can happen when (1) $\dot{e} = 0$ and $\dot{u} \neq 0$, (2) $\dot{e} > 0$ and $\dot{u} < 0$, or (3) $\dot{e} < 0$ and $\dot{u} > 0$. In all those cases, the fractional-order HIGS switches to its lower gain-mode, where its state $x_h$ is kept at zero, guaranteeing continuity of its output signal.
    
\begin{figure}
		\centering
		\begin{subfigure}[b]{0.5\textwidth}
			\centering
			\includegraphics[scale=0.44,trim=20 20 20 4,clip]{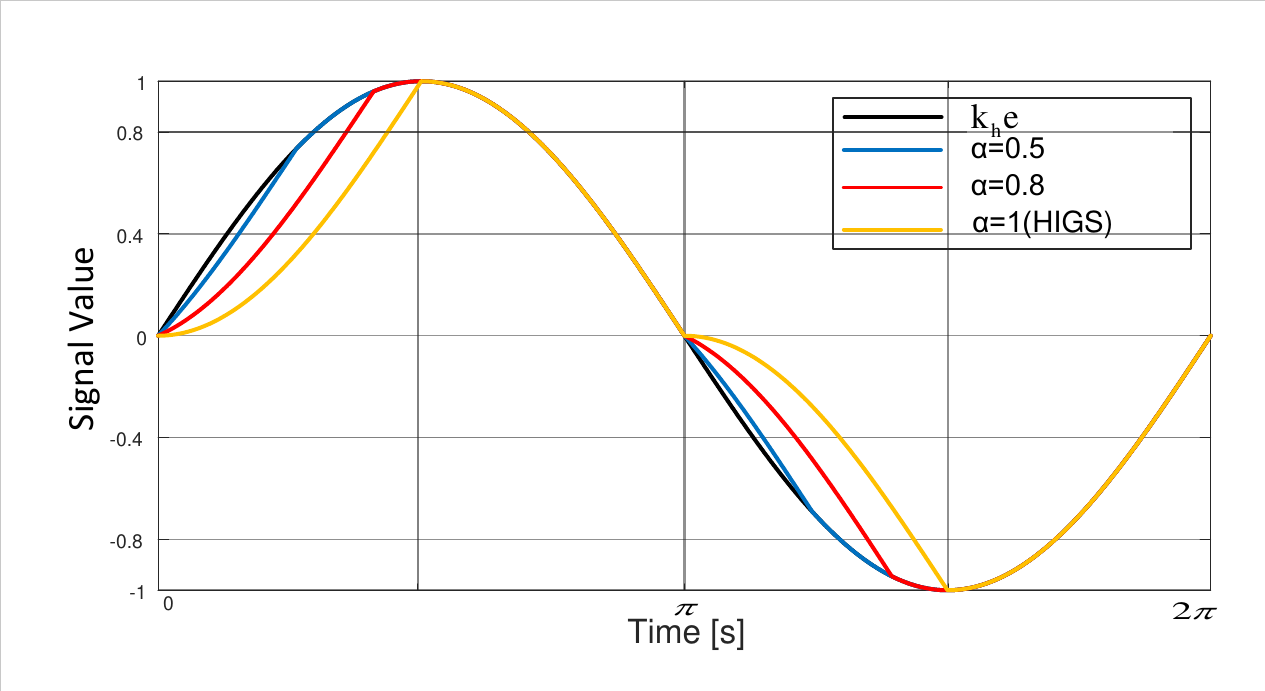}
			\caption{}
			\label{fig:single}
		\end{subfigure}%
		
		~ %add desired spacing between images, e. g. ~, \quad, \qquad etc.
		\begin{subfigure}[b]{0.5\textwidth}
			\centering
			\includegraphics[scale=0.47,trim=22 4 2 4,clip]{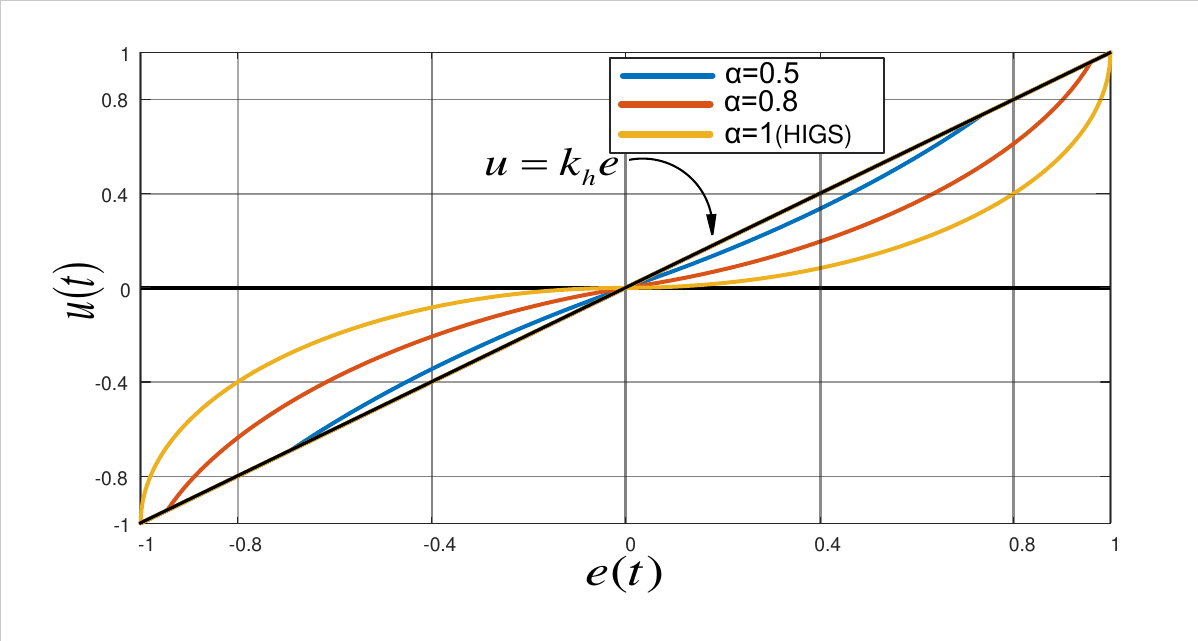}
			\caption{}
			\label{fig:sectore}
		\end{subfigure}
		\setlength{\belowcaptionskip}{-15pt}
		\setlength{\abovecaptionskip}{-10pt}
		\caption{\centering Time domain response of fractional-order HIGS\qquad a) Sinusoidal input response. b) $e-u$ plane.  }
		\label{fig:Fractional}
	\end{figure}

The output of the fractional-order HIGS for a sinusoidal input $e(t)=\sin(t)$ is depicted in Fig. \ref{fig:single}. It can be seen that by decreasing $\alpha$ from 1 to 0 with fixed parameters $\omega_h=1$ and $ k_h=1 $, the output gradually goes from the behaviour of a HIGS to that of a proportional gain. Also, given the importance of HIGS being sector-bounded, Fig. \ref{fig:sectore} shows that the output of the fractional-order HIGS is bounded between $0$ and $k_h e$.

\begin{remark} 
    For the integer-order HIGS it is proven that the second condition of $\mathcal{F}_2$ and all conditions of $\mathcal{F}_3$ are never satisfied, such that they can be disregarded \cite{enh}. However, this does in general not hold for the fractional-order HIGS.
\end{remark}
	%\begin{figure}
	%	\centering
	%	\includegraphics[scale=0.6,trim=4 4 4 4,clip]{Figs/FrHIGSTD.pdf}
	%	\caption{\centering Time domain response of fractional HIGS to a sine wave with various amount of parameter $\alpha$.}
	%	\label{fig.Fr HIGS TD}
	%\end{figure}
	\subsection{Describing function analysis for fractional-order HIGS}
\begin{figure}
		\centering
		\includegraphics[scale=0.52,trim=4 4 4 10,clip]{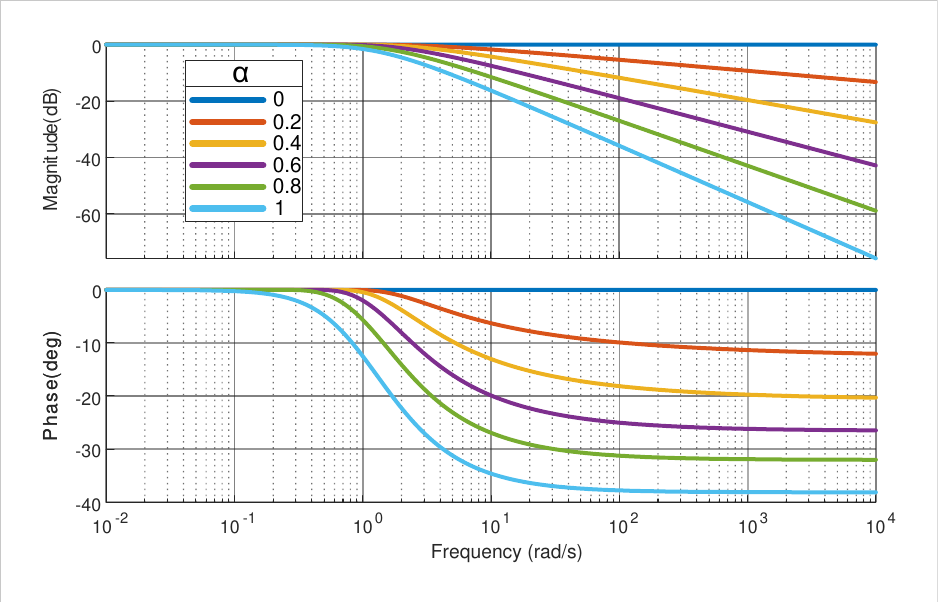}
		\caption{\centering Describing function for fractional-order HIGS with various values of  $\alpha$.}
		\label{fig.Fr HIGS freq}
	\end{figure}

	To derive the describing function of the fractional-order HIGS, first, the time domain output signal of the fractional-order HIGS with input  $e=\hat{e}\sin(\omega t)$ is obtained as:
% 	\begin{equation} 
% 		u(t) = \begin{cases}
% 			\label{eq.Fr HIGS TD}
% 			\omega_h \omega^{-\alpha} \hat{e} \big(\sin(\omega t-\frac{\pi \alpha}{2})+\sin(\frac{\pi \alpha}{2})\big)     & 0\leq t<\frac{\gamma}{\omega} \\
% 			k_h \hat{e} \sin(\omega t)      & \frac{\gamma}{\omega}\leq t<\frac{\pi}{\omega} \\
% 			\omega_h \omega^{-\alpha} \hat{e} \big(\sin(\omega t-\frac{\pi \alpha}{2})-\sin(\frac{\pi \alpha}{2})\big)     & \frac{\pi}{\omega}\leq t<\frac{\gamma +\pi}{\omega} \\
% 			k_h \hat{e} \sin(\omega t).      & \frac{\gamma+\pi}{\omega}\leq t<\frac{2\pi}{\omega} \\
% 		\end{cases}
% 	\end{equation}
\begin{equation} 
		u(t) = \begin{cases}
			\label{eq.Fr HIGS TD}
			\frac{\omega_h \hat{e}}{\omega^\alpha} \big( \sin(\tau-\frac{\pi \alpha}{2}) + \sin(\frac{\pi \alpha}{2}) \big), & \text{if} \hspace{3pt} 0 \leq \tau < \gamma, \\
			k_h \hat{e} \sin(\tau), & \text{if} \hspace{3pt} \gamma \leq \tau < \pi, \\
			\frac{\omega_h \hat{e}}{\omega^\alpha} \big(\sin(\tau-\frac{\pi \alpha}{2})-\sin(\frac{\pi \alpha}{2})\big), & \text{if} \hspace{3pt} \pi \leq \tau < \gamma +\pi, \\
			k_h \hat{e} \sin(\tau), & \text{if} \hspace{3pt} \gamma+\pi \leq \tau < 2\pi, \\
		\end{cases}
	\end{equation}
	where $\tau = \omega t$. For the time intervals $0 \leq \tau < \gamma$ and $\pi \leq \tau < \gamma + \pi$, when the HIGS is in integrator-mode, the output is derived using \eqref{Frsin}. Switching from integrator- to gain-mode happens at $\tau = \gamma$. This time-instant is equal to the moment when the fractional-order HIGS reaches the sector-bound, as given by
	\begin{equation}
		\label{eq:gamma calc}
		\frac{\omega_h\hat{e}}{\omega^{\alpha}} \big(\sin(\gamma -\frac{\pi \alpha}{2})+\sin(\frac{\pi \alpha}{2})\big)=k_h\hat{e} \sin(\gamma).\\
	\end{equation}
	By solving for $\gamma$ we have
	\begin{equation}\label{eq:gamma}
		\gamma  = 2\arctan\left(\frac{\frac{k_h\omega^\alpha}{\omega_h}-\cos(\frac{\pi\alpha}{2})}{\sin(\frac{\pi\alpha}{2})} \right).
	\end{equation}
	The Fourier expansion of  $u(t)$, is given by
	\begin{equation}
		u(t)=\frac{a_0}{2}+\sum_{n=1}^{\infty}(a_n \cos(n\omega t)+b_n \sin(n\omega t))
		\label{Fourier},
	\end{equation}
where Fourier coefficients $a_n$ and $b_n$ are described as
	\begin{subequations}
		\label{eq:anbn}
		\begin{align}
			a_n=\frac{1}{T}\int_{0}^{T}u(t)\cos(n\omega t) \mathrm{d}t, \\
			b_n=\frac{1}{T}\int_{0}^{T}u(t)\sin(n\omega t) \mathrm{d}t,
		\end{align}
	\end{subequations}
	respectively. Here, $n \in \mathbb{Z}^+$ is the order of the harmonic and $T \in \mathbb{R}^+$ is the period of signal $u(t)$. By substituting \eqref{eq.Fr HIGS TD} in \eqref{eq:anbn} and setting $n=1$ we have
	\begin{subequations}
		\label{a1b1}
		\begin{align}
			&a_1= \frac{\omega_h\hat{e}}{4\pi\omega^{\alpha}}\bigg[\cos(\frac{\pi \alpha}{2})-\cos(2\gamma-\frac{\pi \alpha}{2})-2\gamma \sin(\frac{\pi \alpha}{2}) \nonumber  \\
			&+2\cos(\gamma-\frac{\pi \alpha}{2})-2\cos(\gamma+\frac{\pi \alpha}{2})\bigg] \nonumber \\
			&+\frac{k_h\hat{e}}{4\pi}[\cos(2\gamma -1)], \quad \\
			&b_1=\frac{\omega_h\hat{e}}{4\pi\omega^{\alpha}}\bigg[2\gamma \cos(\frac{\pi \alpha}{2})-\sin(2\gamma -\frac{\pi \alpha}{2})+3\sin(\frac{\pi \alpha}{2}) \nonumber \\
			&-2\sin(\gamma + \frac{\pi \alpha}{2})+2\sin(\gamma-\frac{\pi \alpha}{2})\bigg] \nonumber \\
			&+\frac{k_h\hat{e}}{4\pi}[2\pi - 2\gamma+\sin(2\gamma)].
		\end{align}
	\end{subequations}
	According to \cite{khalil}, the describing function is given by
	\begin{equation}
		\mathscr{D}(\omega , \hat{e})=\frac{b_1 + ja_1}{\hat{e}}.
		\label{Df}
	\end{equation}
	Fig. \ref{fig.Fr HIGS freq} shows the describing function of fractional-order HIGS obtained from \eqref{a1b1} and \eqref{Df}. By substituting $\alpha=1$ in \eqref{a1b1} the describing function is equal to that of HIGS in \eqref{HDF}. We can see that by changing parameter $\alpha$, the output phase can vary between $-38^{\circ}$ and $0^{\circ}$. It represents the behavior of the system represented in \eqref{eq.FrHIGSSS}. Also, it verifies that by setting $\alpha$ to zero the system behaves as a proportional gain.
	\begin{figure}
		\centering
		\begin{subfigure}[b]{0.3\textwidth}
			\centering
			\includegraphics[scale=0.4,trim=12 12 12 4,clip]{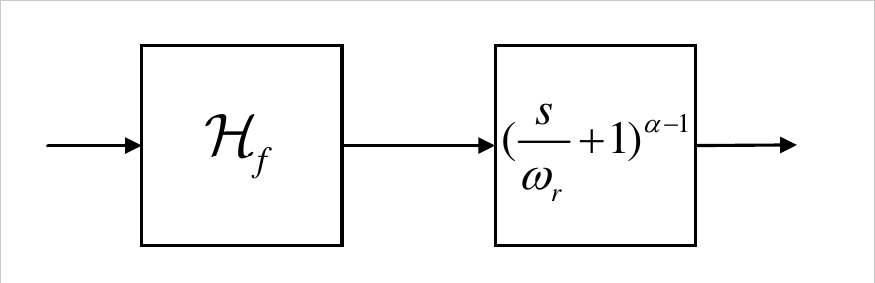}
			\caption{}
			\label{fig:ArchA}
		\end{subfigure}%
		
		~ %add desired spacing between images, e. g. ~, \quad, \qquad etc.
		\begin{subfigure}[b]{0.3\textwidth}
			\centering
			\includegraphics[scale=0.4,trim=4 90 1 4,clip]{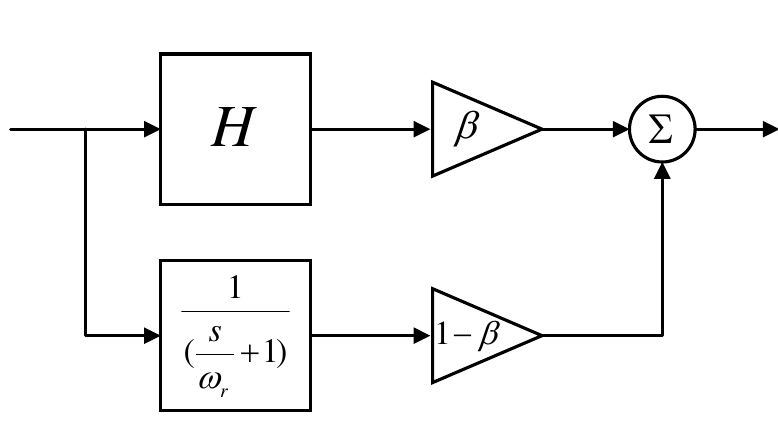}
			\caption{}
			\label{fig:ArchB}
		\end{subfigure}
		\setlength{\belowcaptionskip}{-15pt}
		\caption{\centering Two block diagrams for generalization of HIGS\qquad \quad \quad a) Architecture $a$ ($\mathscr{H}$),\quad  b) Architecture $b$.}
		\label{fig:Architecture}
	\end{figure}
	\subsection{Architecture design for generalization of HIGS}
\begin{figure}
		\centering
		\includegraphics[scale=0.48,trim=4 4 4 4,clip]{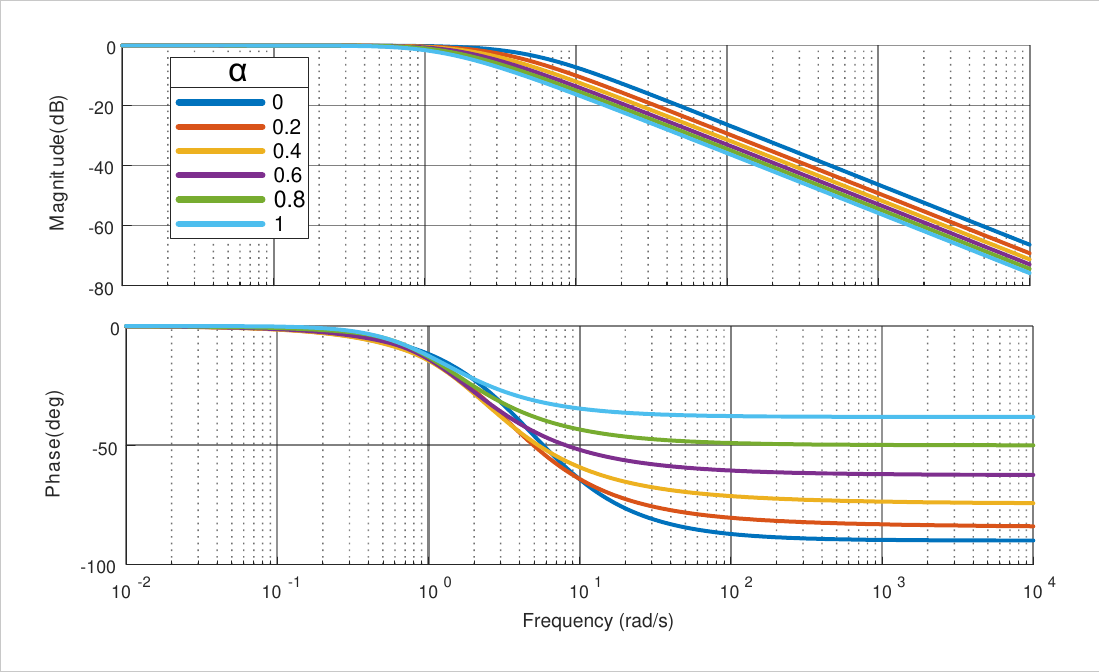}
		\setlength{\belowcaptionskip}{-15pt}
		\caption{\centering Describing function of generalized HIGS ($\mathscr{H}$) with the fractional-order integrator approach.}
		\label{fig.FrHIGS Imp}
	\end{figure}

\begin{figure}
		\centering
		\begin{subfigure}[b]{0.5\textwidth}
			\centering
			\includegraphics[width=8.5cm,height=5cm,trim=20 5 5 8,clip]{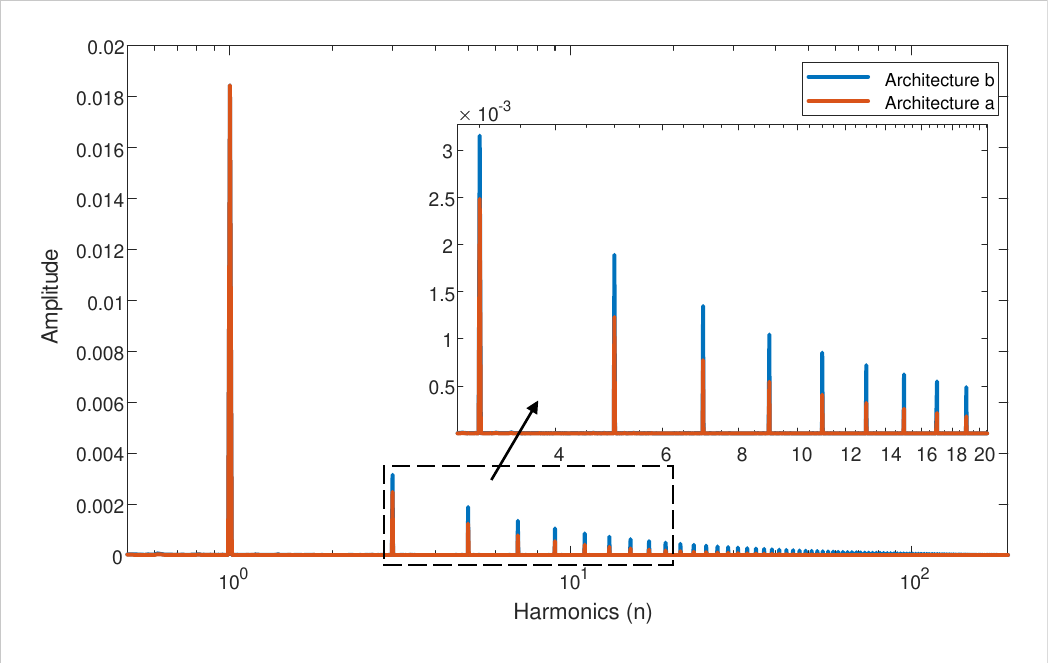}
			%\vspace{8mm}%
			\caption{}
			\label{fig:harmonics}
		\end{subfigure}%
		
		~ %add desired spacing between images, e. g. ~, \quad, \qquad etc.
		\begin{subfigure}[b]{0.5\textwidth}
			\centering
			\includegraphics[width=8.5cm,height=5cm,trim=20 4 5 4,clip]{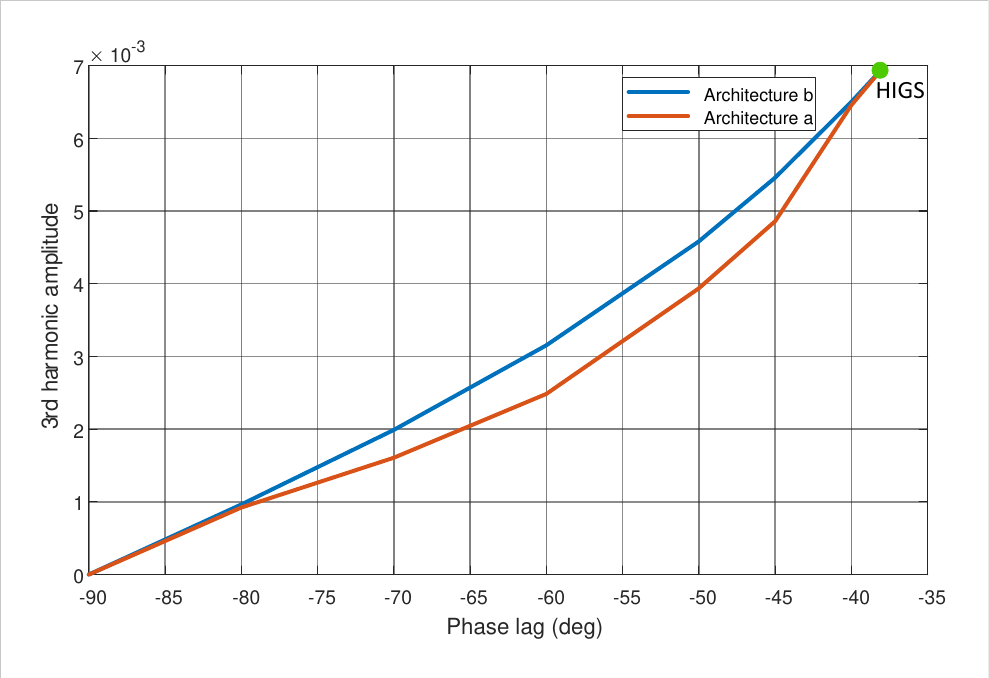}
			\caption{}
			\label{fig:3rd}
		\end{subfigure}
		\caption{\centering a) Higher-order harmonics for Architecture $a$ and $b$ with $\alpha=0.68$ and $\beta=0.5$, b) magnitude of third harmonic for both architectures, obtained by varying $\alpha$ and $\beta$.}
		\label{fig:harmonics2}
	\end{figure}

	In this subsection, we utilize the fractional-order HIGS to construct a generalized HIGS ($\mathscr{H}$) where the parameter $\alpha$ can reproduce a linear low pass filter or HIGS if $\alpha$ is chosen to be 0 or 1, respectively. The architecture of the proposed generalized HIGS is shown in Fig. \ref{fig:ArchA}. Architecture $a$ utilizes a fractional-order HIGS as described in \eqref{eq.FrHIGSSS} and a complementary linear filter to retrieve magnitude-characteristics similar to those of a linear low-pass filter. This complementary filter has the same cut-off frequency ($\omega_r$) as the $\mathscr{H}_f$.
	  Any $\alpha$ between 0 and 1 generalizes the HIGS such that the phase lag of the filter can be varied between $-90^\circ$ and $-38^\circ$, as shown in Fig. \ref{fig.FrHIGS Imp}. With this architecture, the gain of the describing function can be unchanged with respect to the variation of $\alpha$ if each output has a same cut-off frequency ($\omega_r$). A similar analysis can be done for architecture $b$, which is shown in Fig. \ref{fig:ArchB}. Architecture $b$ is inspired by the so-called PI+CI structure in reset control \cite{banos2012reset}. This structure consists of a linear low-pass filter (LPF) and a parallel HIGS element to construct the generalized HIGS. The value of $\beta$ shows the percentage of utilization of each element. Setting this value to 0 represents a linear filter, while the value of 1 results in a HIGS. Any value between 0 and 1 adapts the phase lag of the filter from $-90^\circ$ to $-38^\circ$.
	Although both architectures in Fig. \ref{fig:Architecture} generalize HIGS, architecture $a$ is advantageous over architecture $b$. Since it utilizes a linear low-pass filter after the nonlinear element, it attenuates the higher-order harmonics by $1/n^{(1-\alpha)}$. However, the higher-order harmonics are reduced by a constant value of $1-\beta$ in architecture $b$.\\ To visually illustrate this, we have compared the Fast Fourier transform of both architectures to achieve a generalized HIGS with a phase lag of $-57^{\circ}$. To achieve this phase lag we set $\alpha=0.68$ and $\beta=0.5$ at $\omega=100 \frac{rad}{sec}$. As can be seen from Fig. \ref{fig:harmonics}, the magnitude of the higher-order harmonics for architecture $a$ are smaller than those for architecture $b$, and the magnitudes are decreasing for an increasing $n$. Also, in Fig. \ref{fig:3rd} it is shown that by changing parameter $\alpha$, the magnitude of the third harmonic of architecture $a$  is always less than with architecture $b$. A similar trend can be expected for the other higher-order harmonics.
	\section{Illustrative example}
	In this section, in order to illustrate the time response of generalized HIGS, we control a single mass (double integrator) system with a proportional-integral-derivative (PID) controller. Here, we replace the linear integrator with an integrator made by fractional-order generalized HIGS.
	\subsection{Controller design}
	The PID controller ($C_{pid}$) consists of a lead-lag compensator in series connection to a PI and a proportional gain $K_p \in \mathbb{R}$, given by transfer function
	\begin{equation}
		C_{pid}(s)=K_p\bigg(\frac{1+\frac{s}{\omega_d}}{1+\frac{s}{\omega_t}}\bigg)\bigg(1+\omega_i \mathcal{I} \bigg),
		\label{Cpid}
	\end{equation}
	where $s \in \mathbb{C}$ denotes the Laplace variable, and $\omega_t \in \mathbb{R}$ and $\omega_d \in \mathbb{R}$ are the starting and ending frequency of the lead-action. Furthermore, $\omega_i \in \mathbb{R}$ represents the cut-off frequency of integrator $\mathcal{I}$, which is given by
	\begin{equation}
		\label{I}
		\mathcal{I}=\mathscr{H}\times(1+\frac{\omega_r}{s}). \qquad \qquad  \\
	\end{equation}
	As shown in section III.C, the term $\mathscr{H}$ in \eqref{I} can vary between linear and nonlinear behavior. Therefore, the nonlinearity of integrator $\mathcal{I}$ can be controlled by parameter $\alpha$. In the PID structure, setting $\alpha$ to zero yields linear integrator behaviour for $\mathcal{I}$, and fully nonlinear behaviour when setting $\alpha$ to one. For $\alpha$ between 0 and 1, integrator $\mathcal{I}$ yields a trade-off between both behaviours.\\
	Without loss of generality, we have chosen the crossover frequency as $\omega_c=200\pi (100Hz)$, and set parameters of the controller as $K_p=\omega_c^2/1.8$, $\omega_d=\omega_c/1.8$, $\omega_t=1.8\omega_c$ and $\omega_i=\omega_c/10$. Furthermore, the parameters of $\mathscr{H}$ are set as $\alpha=[0, \,0.1, \,0.2, \,0.4, \,0.7, \,0.9, \,1$], $\omega_h=[ -, \:0.874, \:0.794, \:0.716, \:0.743, \:0.874, \:1$], and $k_h=1$ to have a same cut-off frequency ($\omega_{cf}=1$) as $H$.
	\subsection{Architecture design}
\begin{figure}
		\centering
		\includegraphics[scale=0.4,trim=4 4 4 4,clip]{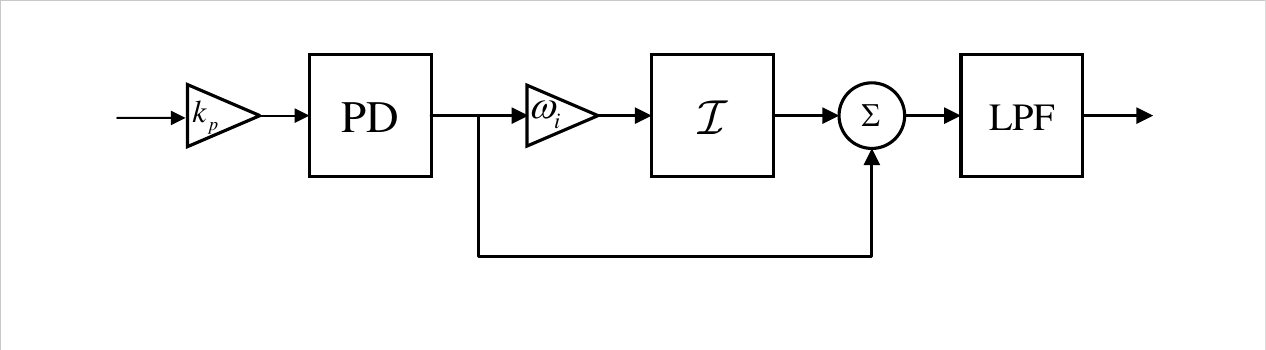}
		\setlength{\belowcaptionskip}{-15pt}
		\setlength{\abovecaptionskip}{-15pt}
		\caption{\centering Simplified schematic of the PID controller with integrator $\mathcal{I}$, that can be variable between a linear and non-linear integrator.}
		\label{fig.Arch C}
	\end{figure}

	The sequence of elements in reset controllers and HIGS controllers is very important. In \cite{cheng}, it has been shown that for a lead-lag compensator in series with a reset element, the optimal sequence for reducing the magnitude of higher-order harmonics is Lead-Reset-Lag. Therefore, the lead-lag compensator in \eqref{Cpid}, is divided into two elements as below:
	\begin{equation}
		\label{divid}
		\bigg(\frac{1+\frac{s}{\omega_d}}{1+\frac{s}{\omega_t}}\bigg) \implies \underbrace{\bigg({1+\frac{s}{\omega_d}}\bigg)}_{PD} \times \underbrace{\bigg(\frac{1}{1+\frac{s}{\omega_t}}\bigg)}_{LPF}
	\end{equation}
	According to the above decomposition, the block diagram of the controller is depicted in Fig. \ref{fig.Arch C}. Note that in an actual system, the PD term should be implemented in a proper form. \vspace{-4mm}
	\subsection{Results}
\begin{figure}
		\centering
		\includegraphics[scale=0.32, angle =180, trim=4 15 12 4,clip]{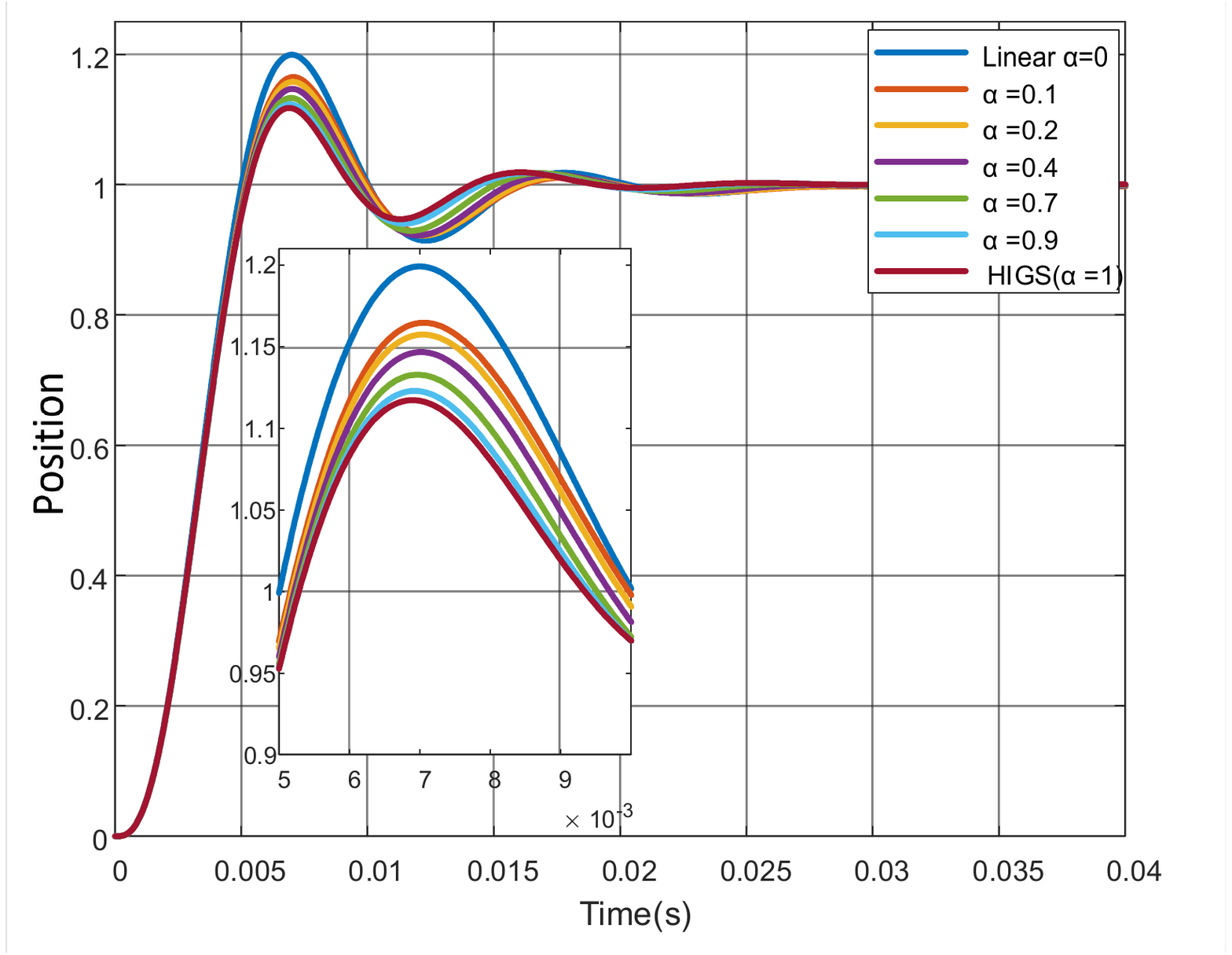}
		\setlength{\belowcaptionskip}{-15pt}
		\caption{\centering Step response of the closed-loop system with a PID controller made by generalized HIGS.}
		\label{fig.VaryAlpha}
	\end{figure}

	In Fig. \ref{fig.VaryAlpha}, the step response of the closed-loop system with controller $C_{pid}$ \eqref{Cpid} is depicted.
	The figure confirms that by changing parameter $\alpha$, the output signal can vary between the responses of linear and nonlinear control systems. By looking at the step response of the system from Fig. \ref{fig.VaryAlpha}, it is clear that by increasing the nonlinearity of the controller, overshoot and settling time are decreased. Therefore, the closed-loop system has better performance in transient response.\\
	In this example, the step response shows the improvement in transient response, where the conventional HIGS performs the best of all generalized HIGS elements. It is not the purpose of this paper to show the superiority of either of these controllers but rather to open a path for more flexible design in the future.\\
	Fractional-order HIGS can be advantageous in improving steady-state response. As can be seen from Fig. \ref{fig:3rd}, the magnitude of the conventional HIGS's third harmonic (the green dot) is the greatest of all possible generalized HIGS elements. It can be problematic, especially for more complex systems with high-frequency modes and control systems where the precision is of concern \cite{karbasizadeh2021fractional}. The generalized HIGS compromises between improving the transient response and reduction of higher-order harmonics for higher precision and reference tracking. 
	\section{Conclusions}
	It is known that HIGS can overcome fundamental limitations of linear control, which leads to better performance without the harmful behavior of reset controllers. Unlike reset control systems, where the reset value can be tuned and its nonlinearity level (phase lag of the describing function) is controllable, the phase lag in HIGS cannot be tuned. In this paper, we proposed a novel fractional-order HIGS that overcomes the aforementioned limitation in HIGS and extends it for more general applications. The describing function was determined for the proposed filter. According to the describing function, it has been shown that the phase lag of generalized HIGS is variable between $-38.15^{\circ}$ and $-90^{\circ}$. Hence by using this system as an integrator, the output can vary between linear and nonlinear behaviors. In the end, the proposed new filter was used in form of a PID to control a double integrator (mass) system. The results validate the generality of generalized HIGS that can be utilized in the future to construct a CgLp element.
	\bibliographystyle{unsrt}
	\bibliography{myrefs}

\end{document}